\documentstyle[aps,prl,psfig,multicol]{revtex}
\begin{document}
\draft
\widetext
\title{Radiative transfer theory for vacuum fluctuations}
\author{
E. G. Mishchenko$^{1,2}$ and 
C. W. J. Beenakker$^{1}$}
\address{{}$^1$Instituut-Lorentz, Universiteit Leiden, P.O. Box 9506, 
2300 RA Leiden, The Netherlands}
\address{{}$^2$ L. D. Landau Institute for Theoretical Physics,
Russian Academy of Sciences, Kosygin 2, Moscow 117334, Russia}

\maketitle
\begin{abstract}
A semiclassical kinetic theory
is presented for the fluctuating photon flux
emitted by a disordered medium in thermal equilibrium.
The kinetic equation is the optical analog of the
Boltzmann-Langevin equation for electrons.
Vacuum fluctuations of the electromagnetic field
provide a new source of fluctuations in the photon flux,
over and above the fluctuations due to scattering.
The kinetic theory in the diffusion
approximation is applied to the super-Poissonian noise
due to photon bunching and to the excess noise due to beating
of incident radiation with the vacuum fluctuations.
\end{abstract}

\begin{multicols}{2}
\par The theory of radiative transfer was developed by Chandrasekhar
\cite{Cha} and Sobolev \cite{Sob} to describe the scattering 
and absorption
of electromagnetic radiation by interstellar matter. It has become
widely used in the study of wave propagation in random media, with
applications in medical imaging and seismic exploration \cite{NATO}.
The basic equation of radiative transfer theory is a kinetic
equation of the Boltzmann type, that is derived from the Maxwell equations
 by
neglecting interference effects \cite{Ish}. It is a
 reliable approximation whenever the scattering and
absorption lengths are large compared to the wavelength,
which applies to all but the most strongly disordered media.
\par Radiative transfer theory has so far been restricted to classical waves,
excluding purely quantum mechanical effects of vacuum fluctuations. 
This limitation is felt strongly in connection with the recent activity 
on random lasers \cite{CZH}. These are amplifying systems in which the feedback 
is provided by multiple scattering from disorder rather than by mirrors,
so that radiative transfer theory is an appropriate level of description.
However, while {\it stimulated} emission has been incorporated into
this approach a long time ago by Letokhov \cite{Let},
{\it spontaneous}  emission has not. It is the purpose of our work
to remove this limitation, by presenting an extension of 
the radiative transfer equation that includes vacuum fluctuations and 
the associated spontaneous emission of radiation.
\par Our inspiration came from the field of electronic
conduction in disordered metals, where the notion of a fluctuating Boltzmann equation (or Boltzmann-Langevin equation)
has been developed extensively \cite{KSh,GGK,Kog}, following the
original proposal by Kadomtsev \cite{Kad}. In that context the fluctuations originate from random scattering
and they conserve the particle number. This same class of
fluctuations exists also in the optical context considered here,
but with a different correlator because of the difference between boson and fermion statistics. In addition, the photons
have a new class of fluctuations, without particle conservation,
originating from random absorption and emission events.
Vacuum fluctuations are of the second class. We will extend
the radiative transfer theory to include both classes of fluctuations.
To demonstrate the validity of our 
``Boltzmann-Langevin equation for photons'', 
we solve the problem of the excess 
noise from vacuum fluctuations in a waveguide geometry, 
for which
an independent solution is known \cite{PB}.
We then apply it to the 
unsolved problem of the thermal radiation from
a spherical random medium.

\par The basic quantity of the kinetic theory is the
fluctuating distribution function $f_{\bf k} ({\bf r},t)$
of the number of photons per unit cell $(2\pi)^{-3} d{\bf k} d{\bf r}$
in phase space. (For simplicity, we ignore the polarization 
dependence.)
Conventional radiative transfer theory deals with the mean
$\bar{f}_{\bf k} ({\bf r})$, which we assume to be time-independent. 
It
satisfies the Boltzmann equation
\begin{equation}
\label{kin}
c\hat{\bf k} \cdot
\frac{\partial \bar{f}_{\bf k}}{\partial {\bf r}} = 
\sum_{\bf k'}
\bigl( J_{\bf kk'}(\bar{f}) -J_{\bf k'k}(\bar{f}) \bigl)+
I^{+}_{\bf k}(\bar{f})-I^{-}_{\bf k}(\bar{f}).
\end{equation}
[For ease of notation, we write $\sum_{\bf k}$
instead of $(2\pi)^{-3} \int d{\bf k}$,
and $\delta_{\bf k q}$ instead of $(2\pi)^{3} \delta 
({\bf k}-{\bf q})$.]
The left-hand-side is the convection term (with $c$ the velocity of 
light in the medium and $\hat{\bf k}$ a unit vector in
the direction of the wavevector ${\bf k}$). The 
right-hand-side contains gain and loss terms
due to scattering, $J_{\bf kk'}(\bar{f})
=w_{\bf kk'}\bar{f}_{\bf k'}
(1+\bar{f}_{\bf k})$, 
due to amplification, $I^{+}_{\bf k}(\bar{f})=
w_{\bf k}^+ (1+\bar{f}_{\bf k})$, and due to
absorption,  $I^{-}_{\bf k}(\bar{f})=
w_{\bf k}^-\bar{f}_{\bf k}$. 
The scattering rate $w_{\bf kk'}=w_{\bf k'k}$
is elastic 
and symmetric.
The absorption and amplification rates $w_{\bf k}^{\pm}$ are
isotropic (dependent only on $k=\vert {\bf k} \vert$)
and related to each other by the requirement
that the Bose-Einstein function
\begin{equation}
\label{Bose}
f_{\rm eq}(\omega,T)=[\exp{(\hbar\omega/k_{B}T)}-1]^{-1}
\end{equation}
is the equilibrium solution of Eq.\ (\ref{kin}) (at
frequency $\omega = ck$ and temperature $T$). This requirement
fixes the ratio $w_{\bf k}^-/w_{\bf k}^+=\exp{(\hbar\omega/k_{B}T)}$.
The temperature $T$ is positive for an absorbing medium and 
negative for an amplifying medium such as a laser
\cite{temp}.
\par We now extend the radiative transfer equation (\ref{kin})
to include the fluctuations $\delta f=f-\bar{f}$. Following
the line of argument that leads to the Boltzmann-Langevin
equation for electrons \cite{KSh,GGK,Kog,Kad}, we propose the
kinetic equation
\begin{eqnarray}
\label{kin1}
c\hat{\bf k} \cdot
\frac{\partial f_{\bf k}}{\partial {\bf r}} &=& \sum_{\bf k'}
 \bigl( J_{\bf kk'} (f)-J_{\bf k'k}(f) \bigl) \nonumber\\
&&\mbox{} +
I^{+}_{\bf k}(f)-I^{-}_{\bf k}(f)+ 
 {\cal  L}_{\bf k}.
\end{eqnarray}
The argument is that the fluctuating $f$ is propagated,
scattered, absorbed, and amplified in the same way as the mean $\bar{f}$, 
hence the same convection term 
and the same 
kernels $J_{\bf kk'}, I_{\bf k}^{\pm}$
appear in Eqs.\ (\ref{kin}) and (\ref{kin1}).
In addition, Eq.\ (\ref{kin1}) contains a
stochastic source of photons,
\begin{equation}
\label{sto}
 {\cal L}_{\bf k}=
\sum_{\bf k'}
 \bigl( \delta J_{\bf kk'} -\delta J_{\bf k'k} \bigl)
+
\delta I^{+}_{\bf k}-\delta I^{-}_{\bf k},
\end{equation}
consisting of separate contributions from scattering, 
amplification,
and absorption.
This Langevin term has  zero mean, $
 \bar{\cal L}_{\bf k}=0$, and a correlator that
follows from the assumption that the
elementary stochastic processes $\delta J_{\bf kk'}$,
$\delta I^{\pm}_{\bf k}$
have independent Poisson distributions:
\begin{mathletters}
\label{source}
\begin{eqnarray}
&&\overline{\delta J_{\bf kk'}({\bf r},t)
\delta J_{\bf qq'}({\bf r'},t')}
= \Delta\delta_{{\bf k}{\bf q}} 
\delta_{{\bf k'}{\bf q'}} J_{\bf kk'}(\bar{f}),
\\
&&\overline{ \delta I^{\pm}_{\bf k}({\bf r},t)
\delta I^{\pm}_{\bf q}({\bf r'},t')}= \Delta  
\delta_{{\bf k}{\bf q}}  I^{\pm}_{\bf k}
(\bar{f}),
\\
&&\overline{ \delta J_{\bf kk'}({\bf r},t)
\delta I^{\pm}_{\bf q}({\bf r'},t')}=0, ~
\overline{ \delta I^{+}_{\bf k}({\bf r},t)
\delta I^{-}_{\bf q}({\bf r'},t')}=0,
\end{eqnarray}
\end{mathletters}%
where we have abbreviated $\Delta = 
 \delta({\bf r}-{\bf r'})\delta(t-t')
$.
Substitution into Eq.\ (\ref{sto}) gives the
correlator
\begin{eqnarray}
\label{correl}
\overline{ {\cal L}_{\bf k}({\bf r},t)
{\cal L}_{\bf q}({\bf r'},t')}
= \Delta
\Bigl[ 
\delta_{{\bf k}{\bf q}} \sum_{\bf k'} 
\bigl( J_{\bf kk'}(\bar{f})+J_{\bf k'k}(\bar{f}) \bigr)
\Bigr.
\nonumber\\
\Bigl. \mbox{}
- J_{\bf kq}(\bar{f}) - J_{\bf qk}(\bar{f})+
\delta_{{\bf k}{\bf q}}\bigl(I^{+}_{\bf k}(\bar{f})+
I^{-}_{\bf k}(\bar{f}) \bigr) \Bigr].
\end{eqnarray}
Eqs.\ (\ref{kin1}) and (\ref{correl})
constitute the Boltzmann-Langevin
equation for photons.

\par To gain more insight into this kinetic
equation we make the diffusion approximation,
valid if the mean free path is the shortest length scale
in the system
(but still large compared to the wavelength).
 The diffusion approximation
consists
in an expansion with respect to
$\hat{\bf k}$ in spherical harmonics, keeping only the first
two terms:
$f_{\bf k}=f_0+\hat{\bf k}\cdot  {\bf f}_1$, 
${\cal L}_{\bf k}=  {\cal L}_{0}+
\hat{\bf k} \cdot
\mbox{\boldmath$\cal  L$}_{1}$,
where $f_0,  {\bf f}_1, {\cal L}_{0}$, and 
$\mbox{\boldmath$\cal  L$}_{1}$ do not depend on
the direction $\hat{\bf k}$
of the wavevector, but on its magnitude
$k=\omega/c$ only. 
The two terms $f_0$ and ${\bf f}_1$
determine, respectively, the photon
number density $n = \rho 
f_0$ and
flux density ${\bf j}=\case{1}{3} c\rho  {\bf f}_1$, 
where $\rho(\omega)=4\pi \omega^2(2\pi c)^{-3}$
is the density of states. 
Integration of Eq.\ (\ref{kin1})
gives two relations between $n$ and ${\bf j}$, 
\begin{eqnarray}
\label{asym}
&&{\bf j}= - D
\frac{\partial  n}{\partial {\bf r}}
+\case{1}{3}l\rho 
\mbox{\boldmath$\cal  L$}_1, \\
\label{sym}
&&\frac{\partial }{\partial {\bf
r}}\cdot  {\bf j} = D \xi_a^{-2}(\rho f_{\rm eq}-n)
+\rho  {\cal L}_0,
\end{eqnarray}
where the diffusion constant $D=\case{1}{3}c^2\tau$ and
 mean free path $l=c\tau$ are determined by
 the transport scattering rate
$\tau^{-1}= \sum_{\bf k'}
w_{\bf kk'} (1-\hat{\bf k}\cdot \hat{\bf k'})$.
The absorption length $\xi_a$ is
defined by $D \xi_a^{-2}=w^- -w^+$. 
(An amplifying medium has an imaginary
$\xi_a$ and a negative $f_{\rm eq}$.)
In Eq.\ (\ref{asym}) we have neglected terms of order 
$(l/\xi_a)^2$, which are assumed to be $\ll 1$.
\par Both Eqs.\ (\ref{asym}) and
(\ref{sym})
contain a fluctuating source term. These two terms
${\cal L}_0$ and $\mbox{\boldmath$\cal  L$}_1$ have zero mean
and correlators that follow from Eq.\ (\ref{correl}),
\begin{mathletters}
\label{cor}
\begin{eqnarray}
\label{cor1}
&&\overline{{\cal L}_{0}(\omega,{\bf r},t) 
{\cal L}_{0}(\omega',{\bf r'},t') }
= \Delta' \frac{D}{\rho \xi_a^{2}}
(2f_{\rm eq} \bar{f}_0+f_{\rm eq}+\bar{f}_0), \\
\label{cor2}
&&\overline{\mbox{\boldmath$\cal  L$}_{1}(\omega,{\bf r},t) 
\mbox{\boldmath$\cal  L$}_{1}(\omega',{\bf r'},t') } 
= \openone \Delta' \frac{6c}{\rho l}
\bar{f}_{0}(1+\bar{f}_0), \\
\label{cor3}
&&\overline{{\cal L}_{0}(\omega,{\bf r},t) 
{\cal L}_{1}(\omega',{\bf r'},t') }
= \Delta' \frac{D}{\rho \xi_a^{2}}
(2f_{\rm eq} \bar{f}_1+\bar{f}_1),
\end{eqnarray}
\end{mathletters}%
where we have abbreviated $\Delta' =\delta (\omega-\omega')
\delta (t-t') \delta ({\bf r} -{\bf r'})$.
The correlator (\ref{cor2}) differs from the electronic case 
\cite{Nag,JB,SL} by the factor $1+\bar{f}_0$ instead of $1-\bar{f}_0$.
This is the expected difference between boson and fermion statistics.
The correlators (\ref{cor1}) and (\ref{cor3}) have no
electronic counterpart. They describe the statistics of the vacuum 
fluctuations.

\par To demonstrate how the kinetic theory presented above
 works in a specific situation
we consider the propagation through an absorbing or
amplifying disordered waveguide (length $L$).
 The incident radiation is isotropic.
All transmitted radiation is absorbed by a photodetector
(see Fig. 1).
Because of the one-dimensionality of the geometry
we need to consider only the $x-$dependence of ${\bf j}$ and $n$
(we assume a unit cross-sectional area).
The transmitted photon flux $I=\int_0^{\infty} d\omega ~  j(\omega,L,t)$
fluctuates around its time-averaged value,
$I(t) = \bar{I}+\delta I(t)$.
The (zero-frequency) noise power $P=\int_{-\infty}^{\infty} dt ~ \overline {\delta I(t)
\delta I(0)}$ is the correlator of the fluctuating 
flux.
We will compute $P$ by solving the differential equations
(\ref{asym}) and (\ref{sym}) with boundary conditions
$n(\omega,0,t)=n_{\rm in}(\omega,t)$, $n(\omega,L,t)=0$, dictated
by the incident radiation at one end of the waveguide and by the
absorbing photodetector at the other end.
\begin{figure}[h]
  \unitlength 1cm
  \begin{center}
  \begin{picture}(8,2.2)
 \put(-0.3,-4.4){\includegraphics{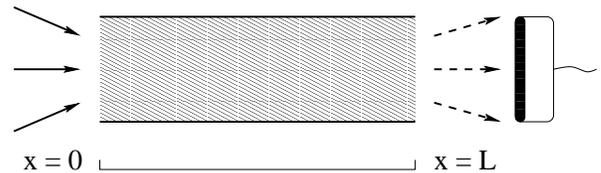}}
  \end{picture}
  \end{center}
  \caption[]{
Isotropic radiation (solid arrows) is incident
on a  waveguide
containing an absorbing or amplifying random medium. The transmitted
radiation (dashed arrows) is absorbed by a photodetector.}
\end{figure}

\par Combining Eqs.\ (\ref{asym}) and (\ref{sym}) we find equations
for the mean and the fluctuations of the photon number,
\begin{eqnarray}
\label{aver}
&&\frac{d^2 \bar{n}}{d x^2}- \frac{\bar{n}}{\xi_a^{2}}
=-\frac{\rho f_{\rm eq}}{\xi_a^{2}},\\
\label{flu}
&&\frac{d^2 \delta n}{d x^2}- \frac{ \delta n}{\xi_a^{2}}
=\frac{\rho}{c}\frac{d {\cal L}_1}{d x} -\frac{\rho {\cal L}_0}{D}.
\end{eqnarray}
The homogeneous differential equation has Green function
\begin{equation}
\label{green}
 G(x,x') = - \xi_a \frac{
\sinh{(x_{<}/\xi_a)} \sinh{(s-x_{>}/\xi_a)}}{\sinh{s}}, 
\end{equation}
where we have defined $s=L/\xi_a$ and
$x_{<}$ ($x_{>}$) is the smallest
(largest) of $x$ and $x'$. 
(In the amplifying system $\xi_a$ is
imaginary so the hyperbolic functions become 
trigonometric functions.)
The inhomogeneous
equations (\ref{aver}) and (\ref{flu}) have the solution
\begin{eqnarray}
\label{prof}
&&\bar{n}(\omega,x)=  \frac{\rho f_{\rm eq}}{\sinh{s}} 
\Bigl( \sinh{s}-\sinh{(x/\xi_a)}
-\sinh{(s-x/\xi_a)} \Bigr)
\nonumber\\ && \; \;\mbox{} +
\bar{n}_{\rm in} (\omega)
\frac{\sinh{(s-x/\xi_a)}}{\sinh{s} },\\
\label{delt}
&& \delta n (\omega,x,t)=\rho \int\limits_0^L dx' ~ G(x,x') \left(
\frac{1}{c}\frac{d {\cal L}_1}{d x'} -\frac{{\cal L}_0}{D} \right)
\nonumber\\ && \; \;\mbox{} +
\delta n_{\rm in} (\omega,t)
\frac{\sinh{(s-x/\xi_a)}}{\sinh{s} }.
\end{eqnarray}
The flux density at the photodetector follows from Eq.\ (\ref{asym})
at $x=L$,
\begin{eqnarray}
\label{curm}
&&\bar{j}(\omega,L)= \frac{D\rho f_{\rm eq}}{\xi_a} \tanh{(s/2)}
+\frac{D\bar{n}_{\rm in}}{\xi_a \sinh{s}}, \\
\label{curf}
&&\delta j(\omega,L,t)= \frac{D\delta n_{\rm in}}{\xi_a \sinh{s}}
+\frac{D\rho}{\sinh{s}}\int\limits_0^L
dx ~ \Bigl(\sinh{(x/\xi_a)} \frac{{\cal L}_0}{D}\nonumber\\ 
&& \; \; \mbox{} +
\cosh{(x/\xi_a)} \frac{{\cal L}_1}{c\xi_a} \Bigr).
\end{eqnarray}
(Notice that the extra term $\propto {\cal L}_1$
in Eq.\ (\ref{asym}) is cancelled by the delta function
in $\partial^2 G/\partial x \partial x'$.)
\par The time-averaged flux $\bar{I}=\bar{I}_{\rm in}+
\bar{I}_{\rm th}$ is the sum of two contributions,
the transmitted incident flux
$\bar{I}_{\rm in}=\int_0^{\infty} d\omega ~ D\bar{n}_{\rm in}
/(\xi_a \sinh{s})$, and the thermal flux
$\bar{I}_{\rm th}=\int_0^{\infty} d\omega (D\rho f_{\rm eq}/\xi_a)
\tanh{(s/2)}$.
The transmitted incident flux per frequency interval
is a fraction $T=4D/(c\xi_a\sinh{s})$
of the incident flux density $j_0=\case{1}{4} c 
\bar{n}_{\rm in}$. A fraction $R=1-4D/(c\xi_a\tanh{s})$
of the incident flux is reflected.
The thermal flux per frequency interval is a 
fraction $1-T-R= (4D/c\xi_a) \tanh{(s/2)}$ 
of the black-body flux density
$j_0=\case{1}{4} c\rho f_{\rm eq}$. This is
Kirchhoff's law of thermal radiation.
\par The noise power $P$ follows from the auto-correlators
of ${\cal L}_0$ and ${\cal L}_1$ [given by Eq.\ (\ref{cor}),
with $\bar{f}_0=\bar{n}/\rho$ from
Eq.\ (\ref{prof})].
The auto-correlator of $\delta n_{\rm in}$
and the cross-correlator of  ${\cal L}_0$ and ${\cal L}_1$
contribute only to order $(l/\xi_a)^2$
and can therefore be neglected.
The noise power
$P=P_{\rm in}+P_{\rm th}+P_{\rm ex}$
is found to consist of three terms, given by
\begin{mathletters}
\label{nois}
\begin{eqnarray}
&& P_{\rm in}=\bar{I}_{\rm in}+ \int\limits_0^{\infty} d\omega ~ 
\frac{D \bar{n}^2_{\rm in}}
{8\rho\xi_a}
\frac{2s\cosh{(2s)}+\sinh{(2s)}-4s}{\sinh^4{s}}, \nonumber\\
\\
\label{noi-th}
&& P_{\rm th}=\bar{I}_{\rm th}+ 
\int\limits_0^{\infty} d\omega ~ \frac{D \rho f^2_{\rm eq}}
{4\xi_a}
\frac{\sinh^2{(s/2)}}{\sinh^4{s}}
\Bigl(8s+4s \cosh{s} \nonumber\\ 
&& \; \; \mbox{} -7\sinh{s}
 -4\sinh{(2s)}+\sinh{(3s)}\Bigr), 
\\
&& P_{\rm ex}= \int\limits_0^{\infty} d\omega ~ \frac{D f_{\rm eq}
\bar{n}_{\rm in}}{2\xi_a}
\frac{\sinh^2{(s/2)}}{\sinh^4{s}}
\Bigl(-6s-4s \cosh{s}\nonumber\\ 
&& \; \;  \mbox{}
+4\sinh{s} +3\sinh{(2s)}\Bigr).
\end{eqnarray}
\end{mathletters}%
The two terms $P_{\rm in}$ and $P_{\rm th}$
describe separately the fluctuations in the transmitted incident 
flux and in the thermal flux.
Both terms are greater than the Poisson noise (the 
mean photon flux $\bar{I}_{\rm th}$, $\bar{I}_{\rm in}$) as a consequence
of photon bunching.
The third term $P_{\rm ex}$ is the excess noise which
in a quantum optical formulation originates from the beating
of the incident radiation with vacuum fluctuations in the
medium \cite{HK}. Here we find this excess noise from
the semiclassical radiative transfer theory. The
expressions for $P_{\rm th}$ and $P_{\rm ex}$
in Eq.\ (\ref{nois}) are the same as those that follow 
from the fully quantum optical treatment \cite{PB,Ben1}.
This is a crucial test of the validity of the semiclassical 
theory. The expression for $P_{\rm in}$
agrees with the quantum optical theory
for the case that the incident
radiation originates from a thermal source
\cite{gauss}.
The case of coherent incident radiation is beyond the reach
of radiative transfer theory. 
\par We envisage a variety of applications for the Boltzmann-Langevin
equation for photons obtained in this paper.
Although we have concentrated here on the waveguide
geometry, in order to be able to compare
with results in the literature, the
calculation of the noise power
in the diffusion approximation can be readily generalized
to arbitrary geometry. 
As an example, we give the noise power of
the thermal
radiation emitted by a sphere (per unit surface area),
\begin{eqnarray}
\label{star}
&&
P_{\rm th}= \bar{I}_{\rm th}+ \int\limits_0^{\infty} 
d\omega~ \frac{2 D\rho f_{\rm eq}^2 s^2}
{\xi_a \sinh^4{s}}
\int\limits_0^s
dz \left(\cosh{z}-\frac{\sinh{z}}{z}\right)^2
\nonumber\\
&& \;\; \mbox{} \times \frac{\sinh^2{z}}{z^2},
\end{eqnarray}
where $s=R/\xi_a$ is the ratio of the
radius $R$ of the sphere and the absorption length
$\xi_a$.
The mean thermal flux is 
given by $\bar{I}_{\rm th} =\int_{0}^{\infty}
 d\omega~ 
D \rho f_{\rm eq} \xi_a^{-1}({\rm cotanh}\, s-1/s)$.
The result for $\bar{I}_{\rm th}$ could have been obtained 
from the conventional radiative transfer theory using
Kirchhoff's law, but the result for $P_{\rm th}$ could not.
\par
A dimensionless measure of the magnitude of the photon flux fluctuations
is the Mandel parameter \cite{MW},
$Q=(P-\bar{I})/\bar{I}$. In a photocount experiment,
counting $n$ photons in a time $t$ with unit
quantum efficiency, the Mandel parameter is obtained
from the mean photocount $\bar{n}$ and the variance ${\rm var}~n$
in the long-time limit: $Q=\lim_{t \rightarrow \infty}
({\rm var}~ n - \bar{n})/\bar{n}$.
We assume a frequency-resolved measurement, so that the 
integrals over frequency in Eqs.\ (\ref{nois}) and (\ref{star})
can be omitted. The Mandel parameter for thermal radiation from a 
waveguide and a sphere is plotted in Fig.\ 2, as a function
of $s$ ($s=L/\xi_a$ for the waveguide and $s=R/\xi_a$
for the sphere). Both the small and large-$s$ behaviour of
$Q$ is geometry independent: $Q= \case{2}{15}
s^2 f_{\rm eq}$ for $s \ll 1$
and $Q=\case{1}{2} f_{\rm eq}$ for $s \gg 1$. The Bose-Einstein
function $f_{\rm eq}(\omega, T)$ is to be evaluated at the detection frequency $\omega$ and temperature $T$ of the medium. The plot in Fig.\ 2 is for
$f_{\rm eq} = 10^{-3}$, typical for optical frequencies at 
$3000~{\rm K}$.
\begin{figure}[h]
  \unitlength 1cm
  \begin{center}
  \begin{picture}(8,5.0)
 \put(-5.8,-11.8){\includegraphics{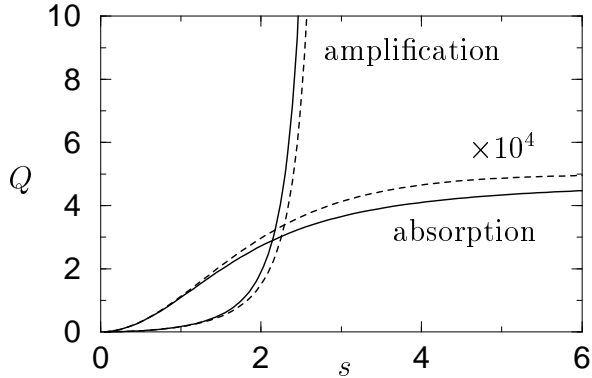}}
  \end{picture}
  \end{center}
  \caption[]{Mandel parameter $Q=(P-\bar{I})/\bar{I}$ for the
thermal radiation 
from an absorbing medium and for the
amplified spontaneous emission from a medium with a complete
population inversion. The solid curves are for the
sphere geometry [Eq.\ (\ref{star})], the dashed curves are for the
waveguide geometry [Eq.\ (\ref{noi-th})]. The parameter
$s$ is the ratio  of the radius of the sphere or
of the length of the waveguide to the absorption or amplification length.
The laser threshold in the amplifying case is at $s=\pi$.
To show both cases in one figure, the $Q$ for the absorbing
medium has been rescaled by a factor $10^4$
(corresponding to $f_{\rm eq}=10^{-3}$).}
\end{figure}
\par Much larger Mandel parameters can be obtained in amplifying systems,
such as a random laser. 
Since complete population inversion corresponds
to $T\rightarrow 0^-$, one has $f_{\rm eq} = -1$ in that case
\cite{temp}.
Eqs.\ (\ref{nois}) and (\ref{star})  apply 
to amplified spontaneous emission below the laser
threshold if one uses
an
imaginary $\xi_a$. 
The absolute value $\vert \xi_a \vert$ is the amplification length, and we 
denote $s=L/\vert \xi_a \vert$ for the waveguide geometry and 
$s=R/\vert \xi_a \vert$ for the sphere.
The laser threshold occurs at $s=\pi$ in both geometries.
We have included in Fig.\ 2 the Mandel parameter for these 
two amplifying systems for the case of complete
population inversion. Again the result is geometry independent
for small $s$, $Q=\case{2}{15} s^2 \vert f_{\rm eq}\vert$ for $s \ll 1$. 
At the laser threshold
($s=\pi$) the Mandel parameter diverges in the
theory considered here.
An important  extension
for future work is to include the non-linearities that become
of crucial importance above
the laser threshold. The simplicity of the radiative transfer theory
developed here makes it a promising tool for the
exploration of the non-linear regime in a random laser.
\par Since radiative transfer theory was originally developed for applications 
in astrophysics, we imagine that the extension to
fluctuations presented here could be useful in that
context as well. 

\par We acknowledge discussions with M. Patra. 
This work was supported by the Dutch
Science Foundation NWO/FOM.

\end{multicols}

\begin{thebibliography}{99}
\bibitem{temp}
The negative temperature of an amplifying medium
is determined by the relative population of the upper and lower levels
of the atomic transition, $N_{\rm lower}/N_{\rm upper}=
\exp{(\hbar\Omega/k_{B}T)}$, where $\Omega$ is the transition
frequency. A complete population inversion corresponds
to $T \rightarrow 0^-$.
\bibitem{gauss}
According to the quantum optical theory of 
Ref.\ \cite{PB}, the integral over frequency in the expression
for $P_{\rm in}$ contains an additional
factor 
$g=\langle a^{\dagger}a^{\dagger}aa \rangle
\langle a^{\dagger} a \rangle^{-2}-1$, where $a$ is the
annihilation operator of the incident
radiation. Eq.\ (\ref{nois})
corresponds to thermal radiation, when $g=1$,
while $g=0$ for coherent radiation. The noise
terms $P_{\rm th}$ and $P_{\rm ex}$ are independent of
$g$.
\end{thebibliography}
\end{document}